\documentclass[aps,pre,twocolumn,showpacs,preprintnumbers,amsmath,amssymb]{revtex4}
\usepackage{graphicx}
\usepackage{bm}

\begin{document}

\preprint{APS/xxx-xxx}

\title{Self-Sustained Collective Oscillation Generated in
an Array of Non-Oscillatory Cells}
\author{Yue Ma}
 \email{dr.mayue@gmail.com}
 \thanks{Corresponding author}
\author{Kenichi Yoshikawa}
 \email{yoshikaw@scphys.kyoto-u.ac.jp}
 \affiliation{Spatio-Temporal Order Project, ICORP,
 Japan Science and Technology Agency (JST) \& \\
 Department of Physics, Graduate School of Science,
 Kyoto University, 606-8502, Japan}

\date{\today}

\begin{abstract}

Oscillations are ubiquitous phenomena in biological systems.
Conventional models of biological periodic oscillations
usually invoke interconnecting transcriptional feedback loops.
Some specific proteins function as transcription factors,
which in turn negatively regulate the expression of the genes
that encode these ``clock proteins''.
These loops may lead to rhythmic changes in gene expression in a cell.
In the case of multi-cellular tissue,
collective oscillation is often due to
synchronization of these cells,
which manifest themselves as autonomous oscillators.
In contrast, we propose here a different scenario
for the occurrence of collective oscillation in a group of 
\emph{non-oscillatory} cells. 
Neither periodic external stimulation nor pacemaker cells with 
intrinsically oscillator are included in present system.
By adopting a spatially inhomogeneous active factor, 
we observe and analyze a coupling-induced oscillation, 
inherent to the phenomenon of 
wave propagation due to intracellular communication.

\end{abstract}

\pacs{A PACS will be appear here}

\maketitle

\section{Introduction}
\label{sec:intro}

Oscillation is ubiquitous in nature, 
not only in physics and chemistry but also biology. 
Biological oscillations can be
observed over a wide range of time- and population-scales, 
from a circadian rhythm of about 24 hours~\cite{Schibler2005Cel} 
to a segmentation clock of less than 2 hours~\cite{Pourquie2003Clock},
from whole-body oscillatory fevers~\cite{Stark2007Immune}
to periodic protein production in a single cell~\cite{Tiana2007Osc}. 
On the other hand, sound theoretical studies have been undergoing since 
long before the observation became possible in molecular level. 
There are many theoretical models to explain these phenomena.
Despite their diversity of biological insights, 
these models share some common points.

Proteins are produced by the transcription and translation
of specific sequences of DNA. 
On the other hand, 
proteins can bind to a transcription promotor on DNA 
and hence suppress or enhance gene expression. 
A transcriptional negative feedback 
loop~\cite{Hirata2002Osc, Hoffmann2002NFKappaB} and a 
delay~\cite{Lewis2003Delay, Chen2002CAS1_Osc} in the inner cellular
gene-protein network are considered to be important elements that 
contribute to the oscillatory expression of DNA and protein production.
From the perspectives of dynamical systems, 
such oscillations are limit cycles that can be generated from 
Hopf bifurcation by choosing an appropriate parameter set 
and initial condition. 
Consequently, in the case of a cell group or
multi-cellular organism with an oscillatory character, 
such as cardiac tissue and a segmentation clock 
in the tail of PSM (Presomitic Mesoderm),
the synchronization of coupled oscillators is often used to explain
the observed collective 
oscillation~\cite{Masamizu2006Rea, Gonze2006Cir, GarciaOjalvo:2004PNAS}.

However, periodic oscillation is only a small part of 
the dynamical behavior of a cell. 
Oscillation may cease if the conditions are changed, 
and most cells tend to settle into a seeming stable state. 
For example, electrical activity in $\beta$-cells exhibits slow  
periodic oscillation at the macro scale of islets of Langerhans, 
while much faster excitability instead of oscillation when 
isolated~\cite{PerezArmendariz:1991p3413, Bertram:2000p3415}. 
In another example, the three proteins (KaiA, KaiB, and KaiC), 
identified as important for the circadian rhythms in 
cyanobacterium \emph{Synechococcus elongates}, 
behave as a bistable toggle switch 
due to a double-negative-feedback loop. 
Oscillation could then arise from the successive switch 
between these two stable steady-states~\cite{Rust:2007p3694, 
Mehra:2006p3698}. 
Moreover, most recent studies also suggested that 
negative transcriptional feedback is not sufficient, 
and in some cases not even necessary, for circadian oscillation. 
Instead, intracellular signaling, 
such as that involving Ca$^{2+}$ and cAMP,
together with transcriptional feedback plays a key role in 
long-term circadian pacemaking~\cite{Neill_Science2008}. 
These evidences raise the possibility that intrinsic oscillatory cells 
are not indispensable in an oscillatory organism. 

In this paper, we study the occurrence of collective oscillation 
from non-oscillatory system. 
In contrast to the conventional mechanism of synchronized oscillators, 
none of the individual cells in our model is intrinsically oscillatory. 
A few studies in the context of mathematics and physics have revealed
the possibility of collective oscillating patterns.
The first example was proposed by Smale~\cite{Smale:1974},
who found that two ``dead'' cells can become ``alive'' 
via diffusive coupling.
More recently, other studies have examined this behavior 
in detail~\cite{Pogromsky:1998ijbc, Pogromsky:1999ijbc}.
In-phase and anti-phase self-sustained oscillation of 
excitable membrane via bulk coupling have been 
observed~\cite{GomezMarin:2007p2974}.
The models considered in these reports have mostly involved 
coupled identical excitable cells with mono-stability.
Some more complicated approaches include, for example, 
using a unidirectional coupling scheme~\cite{In:2003p3688}, 
applying  a periodic stimulation~\cite{yanagita:036226}, 
coupling  the system with an oscillatory 
boundary~\cite{nekhamkina:066224}, 
introducing heterogeneity into excitable 
media~\cite{Cartwright:2000p3406}, 
activity propagating in discrete cellular automata 
model ~\cite{Lewis:2000p3554}, and so forth. 
A commonly used idea is to set isolated cells at 
a subthreshold quiescent state, 
and then push them over into the oscillatory regime to 
generate pacemaker cells by extra force or coupling. 
That means cells are possible to manifest themselves as oscillators.  
However, little attention has been paid to the emergence of oscillation 
in systems that are completely independent of oscillating elements. 
Unlike previous studies, geometrical structure of nullclines of cells 
in our model prevent dynamics from being oscillation.  
There are two ``engines'' in our model to drive the 
self-sustained collective oscillation, 
neither of which is oscillatory pacemaker. 
The one is bistable cell switching between two stable states, 
the other is mono-stable cell with excitability. 
Two engines work cooperatively due to the wave propagation.


On the other hand, in a bistable system,
a stationary front can bifurcate into a pair of fronts
that propagate in opposite directions,
which is known as non-equilibrium Ising-Bloch (NIB) 
bifurcation~\cite{Coullet:1990p631, Hagberg:1994p553}.
Perturbation for the occurrence of NIB bifurcation can be 
induced by local spatial inhomogeneity~\cite{Bode:1997PhyD}.
A more global analysis showed that the NIB point is only part of the story,
and concluded that an unstable wave front is intrinsic to media 
that are spatially inhomogeneous~\cite{Prat:2003p636, Prat:2005p638}.
An unstable wave front may manifest itself as a reflected front,
tango wave~\cite{Li:2003p637}, pacemaker~\cite{Miyazaki:2007p739} and so on.
In this paper, we think about these phenomena 
beyond mathematics and physics,
and extend their application to biological oscillators.

Moreover, although most studies have been performed on
a spatial continuum described by partial differential equations (PDE),
continuum models neglect the effects of cellular 
discreteness~\cite{Shnerb_PNAS2000}.
In fact, from the viewpoint of biology,
the size of cells can not decreased infinitely.
This intrinsic property is difficult to ignore, 
especially at the stage of initial development of an organism,
when the cell size is comparable to that of tissue.
In addition, there are mathematical reasons to explore 
the system dynamics with spatial discretization.
PDE and ODE (ordinary differential equations)
have different theoretical frameworks and produce different results.
Several significant features of discreteness,
such as \emph{wave propagation failure}~\cite{Keener_SIAM1987},
can not occur in a continuum model.
Therefore, in this paper we will consider an array of 
spatially discrete cells, and discuss the impact of discreteness.

\section{Description of the model}

\subsection{One-dimensional cellular array}

In this paper, we consider cells in one-dimensional space. 
Cells are coupled by intracellular signaling molecules,
which flow through channels in a membrane due to concentration difference
or depolarization-mediated flux. 
The intracellular signaling small-molecule can be produced by  
a series of process from some genes functioned as activator, 
and then trigger transcriptional feedback loops of adjacent cells.
We assume that the coupling interaction takes place 
in a diffusion-like manner. 
If we include an inhibitor, 
which can locally repress the expression of activator genes, 
a one-dimensional array of $N$ cells can be described as
\begin{align}
\label{eq:chain1}
    \dot{u}_i &= f(u_i, v_i, \Gamma_i) + \tilde{D}(u_{i-1} + u_{i+1} - 2 u_i) \\
\label{eq:chain2}
    \dot{v}_i &= g(u_i, v_i)
\end{align}
where $u$ and $v$ are concentration of activator and inhibitor, 
respectively, $i\in\{1 \dots N\}$ is the index of the cell in the chain and 
$\tilde{D}$ is the coupling strength of $u$.
$f$ and $g$ are corresponding reaction functions. 
The boundary condition is zero flux, i.e., $u_0=u_1$ and $u_{N}=u_{N+1}$.
Finally, $\Gamma_i$ is an environmental parameter, 
which will be discussed in detail later. 

In this study, we only consider coupling of the activator. 
For most of the models that have been used to study pattern formation, 
diffusion is assumed to occur for every elements. 
Specially, much greater diffusion of the inhibitor is necessary to  
induce Turing instability~\cite{Turing_1952}. 
However, a cells membrane is very selective for passage of substances. 
Complicated intracellular reactions usually take place locally, 
but are triggered by only one or a few specific signaling molecules. 
For example, while the segmentation clock involves the cyclic expression 
of many genes, the crucial pathway for coupling only involves the 
transmembrane receptor Notch1~\cite{Kageyama_2007DevDyn_review}.
Thus, in the context of biology, we only consider coupling 
with the activator, 
and the inhibitor in our model is merely a local state variable. 

\subsection{Active factor}

The development of a multicellular organism begins with a single cell, 
which divides and gives rise to cells with different typologies.
Different cells are organized according to certain secreted chemicals,
called morphogens.
Despite improvements in experimental and theoretical approaches,
the mechanisms of morphogenesis are still unclear.
Usually, morphogens are considered to be produced at specific sites
and diffuse through the organism~\cite{Ibanes_MSB2008}. 
Quite recently, evidence of a ``shuttling-based'' 
mechanism has been presented~\cite{Ben-zvi_Nature2008}.
The key in such models is their ability to define a robust and scaling profile,
usually a concentration gradient, of morphogens.
More broadly, 
we can suppose that some environmental parameters act as morphogens.
The environment in which an organism develops supplies nutrition for growth,
and the intracellular volume in direct contact with the border 
gets more and that deep inside cells gets less.

In this paper, we do not consider any specific chemical substance,
and instead merely suppose that there is a certain factor, 
which we refer to as the \emph{active factor}, 
to obtain information regarding the relation 
between position and cell dynamics. 
The above-mentioned active factors can affect the fate of cells in a
concentration dependent manner~\cite{Wolpert1969Pos, Lewis:2008p3436}. 

\begin{figure}[t!]
\centering
\includegraphics[width=0.9\columnwidth]{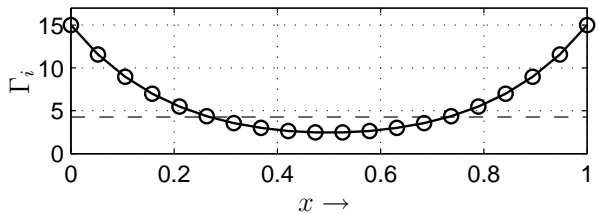}
\caption{Profile of $\Gamma_i$ obtained from Eq.(\ref{eq:Gamma-profile}), 
when $\Gamma_{0} = 15, \xi = 5$.}
\label{fig:Gamma}
\end{figure}

Without losing generality, we assume that the active factor
$\Gamma$ is constant at the boundaries of an organism, 
where the source site of morphogens are usually located.
It diffuses into the organism field with a diffusion constant $D_a$,
and is degraded at rate $\alpha$.
Thus, we have
\begin{equation}
 \label{eq:diffusion-degradation}
  \frac{\partial \Gamma}{\partial t} =
  D_a \frac{\partial^2 \Gamma}{\partial x^2} - \alpha \Gamma .
\end{equation}
Since our model is based on coupled ODEs independent of spatial variation, 
the profile of the active factor satisfies a scaling property.
By normalizing the field size to one, we can get a steady profile
($\partial \Gamma / \partial t = 0$) of $\Gamma$ as
\begin{equation}
 \label{eq:Gamma-profile}
  \Gamma(x) = \frac{\Gamma_0}{e^{-\xi}-e^{\xi}}
  ((e^{-\xi}-1) e^{\xi x}
  - (e^{\xi}-1) e^{-\xi x}) ,
\end{equation}
where $\Gamma_0 = \Gamma(0) = \Gamma(1)$ is the value at two boundaries,
and $\xi = 1/\lambda = \sqrt{\alpha / D_a}$ 
is the inverse of the decay length.
A typical profile of $\Gamma(x)$ is shown in Fig.~\ref{fig:Gamma}.
Circles indicate the value of $\Gamma$ for discrete cells
($N=20$ in the figure) placed uniformly in the scaling field.

\subsection{Bistability}

We assume that cells normally prefer to live in a stable state, 
and cells with a high concentration of active factor 
are capable of switching between two states. 
This kind of bistability is very important and has been observed
in various biological systems~\cite{Angeli2007Det}.  
For example, the expression of the \emph{Dictyostelium} cAMP 
phosphodiesterase gene behave as a bistable switch employing 
intracellular cAMP as a regulator of cell fate~\cite{BRiley061990}, 
the Cdc2 activation system in Xenopus egg extracts is bistable and 
characterized by biochemical hysteresis~\cite{Pomerening:2003p3699}, 
the inducible \emph{lac} operon in \emph{E. coli} shows 
bistability~\cite{Santillan:2007p3830}, and so on.

Usually, bistability arises from positive, 
or double-negative genetic regulation
loops~\cite{Kim2008Motifs, Angeli2007Det, Holt_Nature2008_Positive}.
It was recently suggested that stochastic fluctuation 
plays an important role in the nature of the transition 
between bistable states~\cite{Maamar2007Noise, Losick:2008p3089}.
Moreover, physical regulation of protein production,
which has been much less considered by biochemists, 
also plays an important role in the origin of the bistability. 
It has been observed that discrete transition between folding and 
unfolding states, namely a first-order phase transition, 
can take place in giant DNA~\cite{Luckel:2005FEBS}. 
Similar discrete switch can also occur for 
RNA~\cite{mamasakhlisov_PRE2007},
protein~\cite{Schanda_2007PNAS} 
and other molecules~\cite{Etienne_2002Nature}. 
This discrete transition leads to the ON/OFF switching 
of the production of a specific protein.

\begin{figure}[t!]
\centering
\includegraphics[width=\columnwidth]{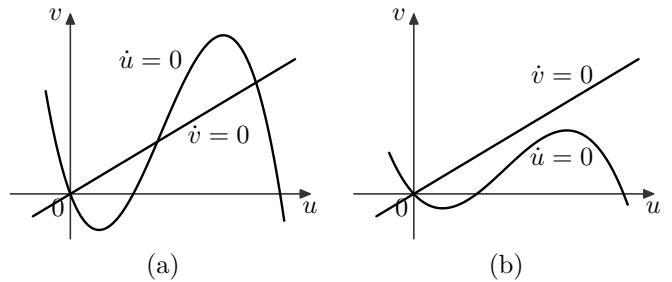}
\caption{Nullcline diagrams in 
bistability (a) and mono-stability (b), respectively.}
\label{fig:onoff}
\end{figure}

\subsection{Model equations}

We describe the dynamical reaction function of each cell 
by using the two-component Fitzhugh-Nagumo equations
\begin{align}
\label{eq:fhn1}
\dot{u} &= f(u, v, \Gamma) = \Gamma u(u - \alpha) (1 - u) - v \\
\label{eq:fhn2}
\dot{v} & = g(u, v) = \epsilon (\beta u - v)
\end{align}
where $u$ is a variable related to the expression level of 
specific activator genes, $v$ is the inhibitor to repress $u$, 
$\epsilon$, which is much smaller than 1, is the slower 
growth factor of inhibitor $v$ and $\Gamma$ is the 
active factor discussed previously. 
Note that the kinetics of inhibitor here is a rather natural 
unit process in many of biochemical reactions. 
Throughout this paper, the following parameters are fixed 
\begin{equation}
\label{eq:alpha-beta-epsilon}
  \alpha = 0.3, \ \beta = 0.5, \ \epsilon = 0.02.
\end{equation}
The Fitzhugh-Nagumo model has been well studied for description 
of excitable behavior in biology. 
Rich nonlinear dynamics can be observed by tuning parameters. 
Specifically, with the above parameters, 
the model is mono-stable at small value of $\Gamma$, 
and will happen a saddle-node bifurcation at $\Gamma=4.08$
and a Hopf bifurcation at $\Gamma=4.27$, 
which leads to bistability.  
Thus, in the case of the spatial profile of 
$\Gamma_i$ as shown in Fig.~\ref{fig:Gamma}, 
only the 8 central cells (7$^{th}$ - 14$^{th}$) are mono-stable, 
while the others are bistable ($\Gamma_6=\Gamma_{15}=4.37$).

Figures~\ref{fig:onoff}(a)\&(b) show the nullclines with bistability 
and mono-stability for when $\Gamma$ is large and small, respectively.
Again, none of the cells show oscillation in the absence of coupling.
More important, from the geometry property of nullclines in the figure, 
no oscillatory condition could be found by moving cubic nullcline 
($\dot{u}=0$) up and down. That is, no pacemaker cells can be 
generated from activator coupling. 

If we substitute Eq.~(\ref{eq:fhn1}\&\ref{eq:fhn2}) into 
Eq.~(\ref{eq:chain1}\&\ref{eq:chain2}), 
we get the system equations used in this paper.
\begin{align}
\label{eq:sys1}
\dot{u}_i &= \Gamma_i u_i (u_i - \alpha) (1 - u_i) - v_i \\ \nonumber
& \ \ \ + \tilde{D} (u_{i-1} + u_{i+1} - 2 u_{i}) \\
\label{eq:sys2}
\dot{v}_i &= \epsilon(\beta u_i - v_i)
\end{align}
When we change the coupling strength $\tilde{D}$, 
we observe the occurrence, variation and disappearance 
of self-sustained collective oscillation in the cell array.

\section{Self-sustained collective oscillation}

\subsection{Normal collective oscillation}

\begin{figure}[t!]
\centering
\includegraphics{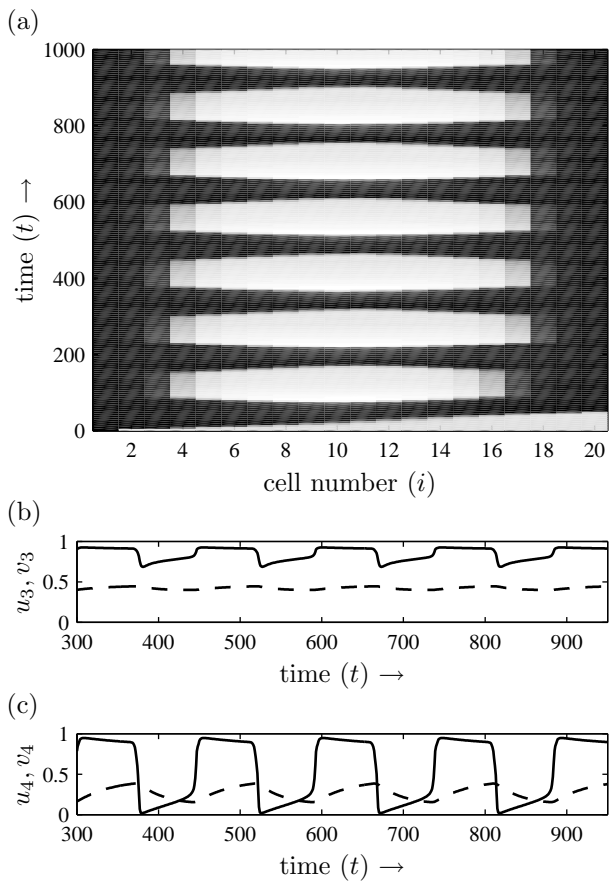}
\caption{Collective oscillation observed in a chain of cells when 
$\tilde{D}=0.7$ in Eqs.~(\ref{eq:sys1}\& \ref{eq:sys2}).
(a) Spatio-temporal plot of the collective oscillation of $u_i$. 
The black and white indicate $u_i=1$ and $u_i=0$, respectively. 
(b, c) Waveforms of $u$ (solid) and $v$ (dashed) in the 3rd and 4th cells.}
\label{fig:normal_osc}
\end{figure}

\begin{figure}[t!]
\centering
\includegraphics{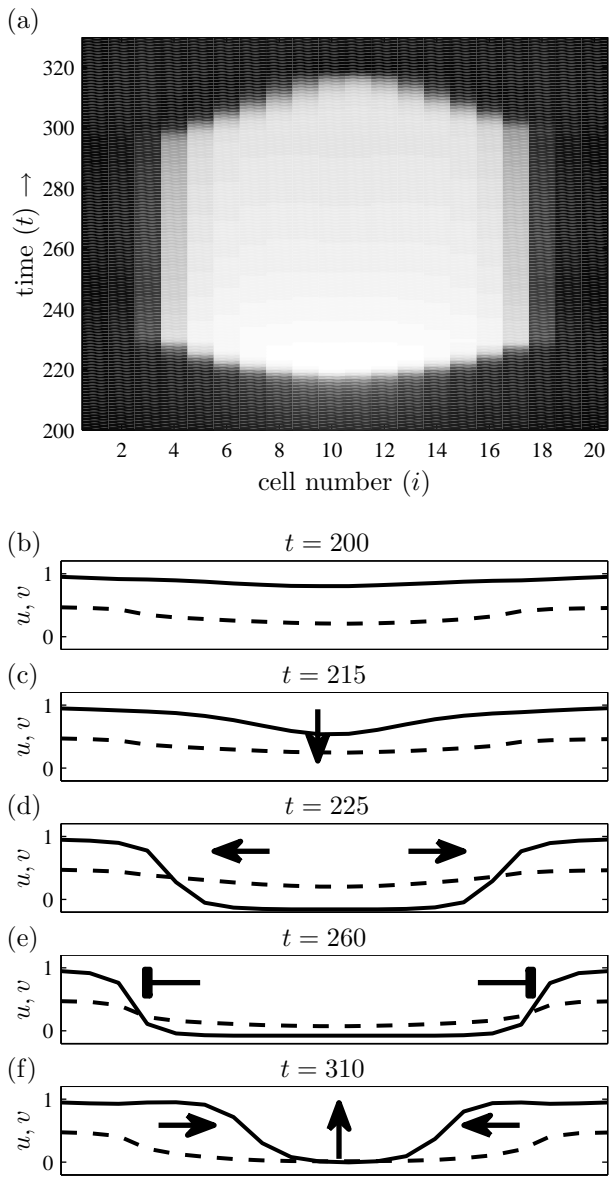}
  \caption{(a) Spatio-temporal diagram of $u_i$ over a single period,
  (b)-(f) Snapshots of $u$ and $v$ at several time points in one period, 
  where the horizontal axis is cell number from 1 to 20.
  This illustrates the change in wave propagation at different stages.
 Solid curves and dashed curves indicate $u$ and $v$,
  respectively. An animation, through which the behavior can be 
  understood more intuitively, is available at \cite{mysite}.}
  \label{fig:wave-1p}
\end{figure}

Figure~\ref{fig:normal_osc}(a) shows a typical oscillation
when the coupling strength $\tilde{D} = 0.7$.
Figure~\ref{fig:wave-1p}(a) shows an enlarged view 
of a single period of oscillation.
As the initial condition, we set the 1st cell as being excited,
since stimulation is  usually input from the border.
Initially, ($0 < t \lesssim 80$),
a traveling wave appears due to excitation at the border.
The traveling front then sweeps over the cell array
and makes all of the cells excited (see Fig.~\ref{fig:wave-1p}(b)).
Although the central cells are also turned ON 
due to the interaction with other cells, 
they can not stay in the excitable state for a long time.
Instead, they soon return to their stable equilibrium 
(see Fig.~\ref{fig:wave-1p}(c)),
and hence generate two counterpropagating wave backs, 
as shown in Fig.~\ref{fig:wave-1p}(d).
These two wave backs propagate outward until the 3$^{rd}$ and 18$^{th}$ 
cells and stop suddenly due to the  
spatial discreteness (see Fig.~\ref{fig:wave-1p}(e)).
The ``wall'' cells do not jump from the ON state to the OFF state
and only exhibit slight oscillation closed to their equilibrium.
As an example, the difference between the 3$^{rd}$ and 4$^{th}$ cells 
is shown in Fig.~\ref{fig:normal_osc}(b, c).
At this critical interface,
the inhibitor $v$ slowly decreases so that
the 4$^{th}$ and 17$^{th}$ cells restore excitability after a while.
The central cells can then be excited again by the pair of 
reflecting wave fronts, as shown in Fig.~\ref{fig:wave-1p}(f).
Pushed by the wave, the central cells will be excited again.
This process repeats and causes the collective oscillation inside
multi-cell tissue without oscillatory cells.

\subsection{Stationary state before birth of oscillation}

\begin{figure}[t!]
\centering
\includegraphics{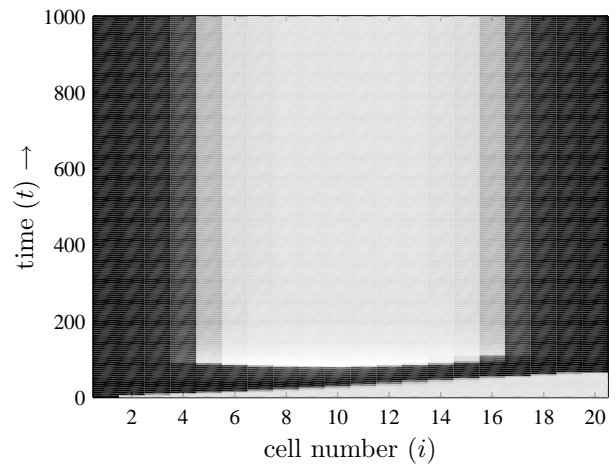}
\caption{Spatio-temporal diagram of $u_i$ in a stationary state.
Wave propagation stops and no oscillation occurs
in the case of weak coupling $\tilde{D}=0.47$.}
\label{fig:beforeosc}
\end{figure}

\begin{figure}[t!]
\centering
\includegraphics{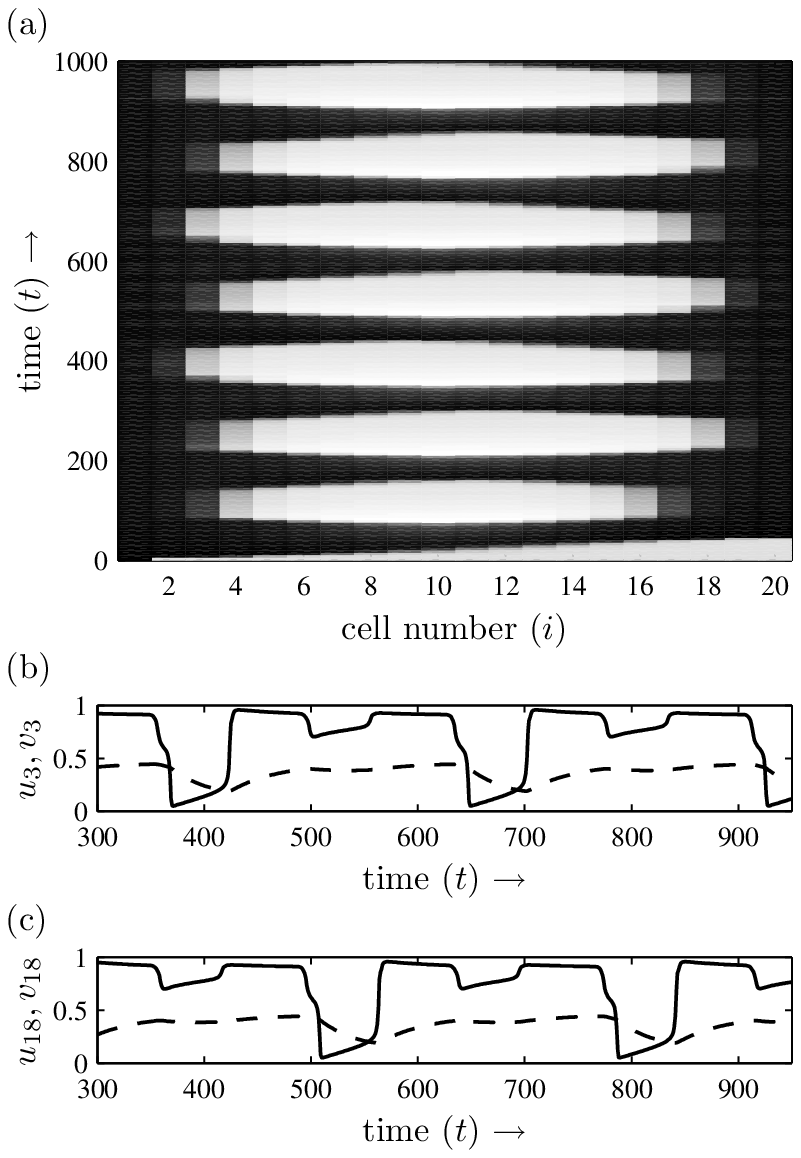}
 \caption{Anti-phase mode in period doubling produces collective 
 oscillation with a periodic position shift when $\tilde{D}=0.8$.
 (a) Spatio-temporal diagram of $u_i$.
 The grayscale black and white indicate $u_i=1$ and $u_i=0$, respectively.
 (b) and (c) are waveform diagrams of the 3rd and 18th cells. 
 Activator $u$ and inhibitor $v$ are shown in solid and dashed curves, 
 respectively.}
 \label{fig:antiphase_osc}
\end{figure}

\begin{figure}[t!]
\centering
\includegraphics{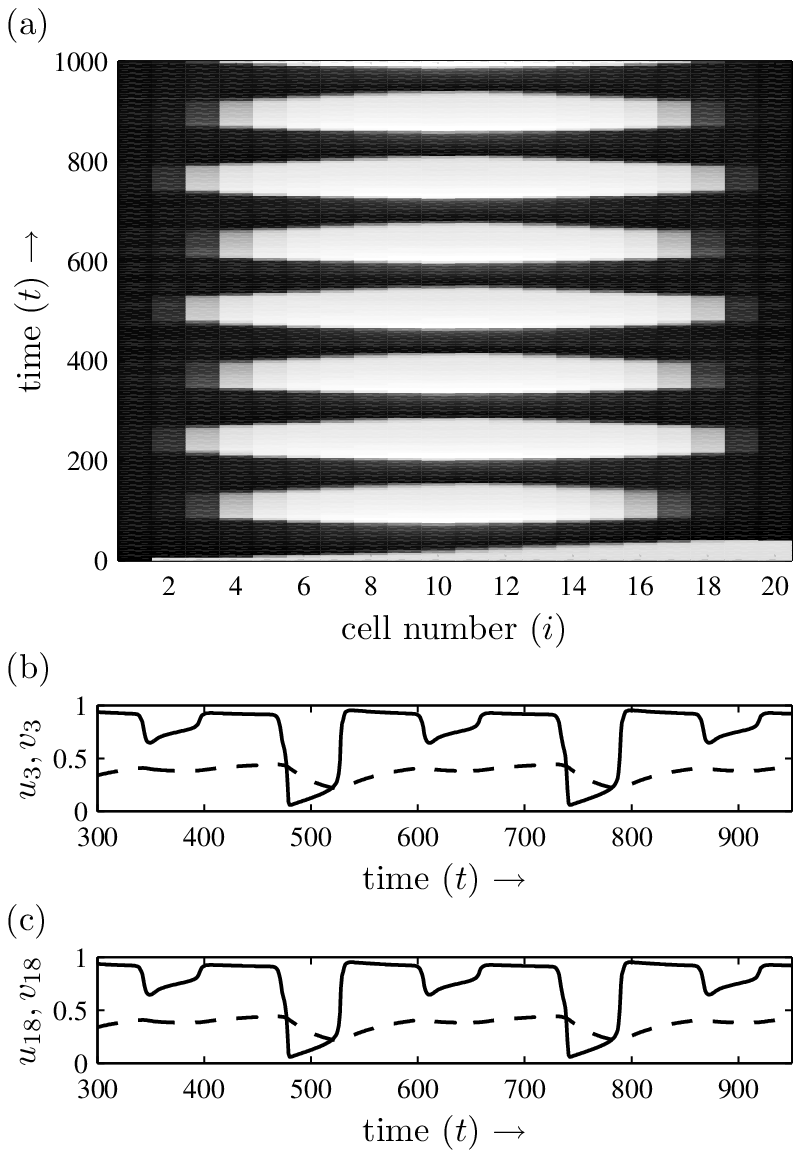}
 \caption{In-phase mode in period doubling produces collective 
 oscillation with a change in the periodic population when $\tilde{D}=0.9$.
 (a) Spatio-temporal diagram of $u_i$. 
 The grayscale black and white indicate $u_i=1$ and $u_i=0$, respectively.
 (b) and (c) are waveform diagrams of the 3$^{rd}$ and 18$^{th}$ cells.
 Activator $u$ and inhibitor $v$ are shown in solid and dashed curves, 
 respectively.}
 \label{fig:inphase_osc}
\end{figure}

The above collective oscillation can be observed 
when the coupling strength is larger than a threshold, 
below which wave backs (see Fig.~\ref{fig:wave-1p}(d)) fail to reflect, 
and the state in Fig.~\ref{fig:wave-1p}(e) is maintained. 
Figure~\ref{fig:beforeosc} shows a spatio-temporal diagram, 
where the central cells stay silent while excited bands appear
close to the two borders.

Note that this phenomenon could not take place in a continuum counterpart.
The existence of a coupling strength threshold under which 
\emph{wave propagation failure} occurs is unique to 
a spatially discrete system.
In addition, there is another threshold, which is even smaller,
for which the \emph{wave front} stops propagating.
In this case, the excited signal at the border fails to propagate forward, 
but we would like to postpone this interesting phenomenon on another paper, 
since it is less related to present work.

\subsection{In-phase and anti-phase period doubling oscillation
closed to the boundary}
With an increase in the coupling strength $\tilde{D}$,
the characteristics of oscillation can be changed.
Figure~\ref{fig:antiphase_osc} shows that 
the position of oscillation periodically shifts.
The 3$^{rd}$ and 18$^{th}$ cells oscillate with a nearly doubled period, 
in anti-phase (Fig.~\ref{fig:antiphase_osc}(b,c)).
Globally, tissue oscillates in two groups with the same
cell populations but different positions: 
No.~3-No.~17 (15 cells) and No.~4-No.~18 (15 cells), respectively.

Interestingly, by slightly increasing the coupling strength $\tilde{D}$,
say to $\tilde{D}=0.9$, we found a different type of period doubling,
as shown in Fig.~\ref{fig:inphase_osc}.
For comparison with the case of $\tilde{D}=0.8$,
although the critical interface between ON and OFF shifts periodically
as in Fig.~\ref{fig:antiphase_osc},
there is no phase difference between the 3$^{rd}$ and 18$^{th}$ cells.
As is clearly shown in their waveform (Fig.~\ref{fig:inphase_osc}(b,c)),
these two boundary cells oscillate in-phase, 
instead of anti-phase (Fig.~\ref{fig:antiphase_osc}(b,c)).
Therefore, in the present condition,
a periodic change does not take place in the position of oscillation.
Instead, the population of oscillating cells changes.
More precisely, tissue oscillates in two groups:
No.~3-No.~18 (16 cells) and No.~4-No.~17 (14 cells), respectively.

Moreover, by setting the initial condition of the cells identically, 
i.e., all in the ON state at $t=0$,
we found checked that same symmetric collective oscillation
can also occur in the case of $\tilde{D}=0.8$.
Therefore, we conclude that these two types of oscillation are caused 
by the same bifurcation.
Because the wavefront propagates faster with larger $\tilde{D}$,
a larger coupling strength can reduce the time lag between 
the two boundary cells being stimulated.
If the time lag is smaller, 
the two boundaries converge to in-phase oscillation. 
On the other hand, if the time lag is large, 
they will exhibit anti-phase oscillation.

\subsection{Oscillation death}

\begin{figure}[t!]
 \centering
  \includegraphics[width=\columnwidth]{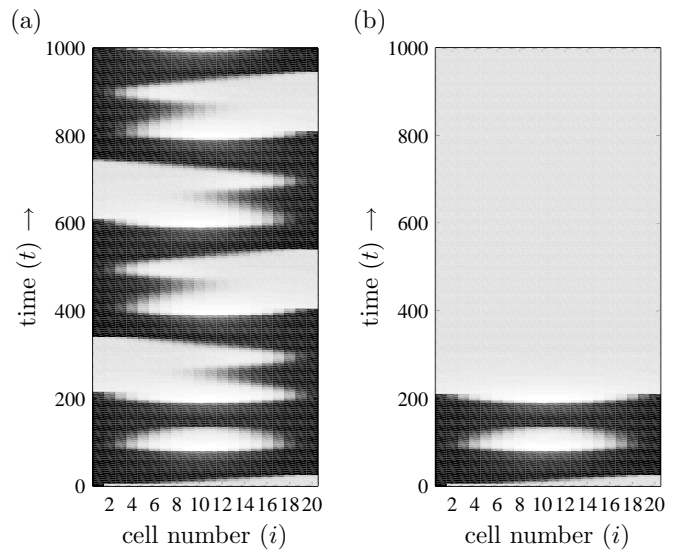}
 \caption{Spatio-temporal diagram of $u_i$.
   Black and white indicate $u_i = 1$ and $u_i = 0$, respectively.
   (a) $\tilde{D}=2.53$, oscillation starts to collapse;  
   (b) $\tilde{D}=2.6$, oscillation ceases after one cycle.}
 \label{fig:osc_death}
\end{figure}

With a increase in coupling strength $\tilde{D}$,
we observed that the change in the periodic position or population stopped, 
and normal oscillation returned. 
In comparison to the case of $\tilde{D}=0.7$,
the total population of oscillating cells increased from 14 (No.~4 to No.~17) 
to 16 (No.~3 to No.~18).

The oscillation suddenly dies when $\tilde{D}$ is as large as 2.6.
Figure~\ref{fig:osc_death}(b) clearly shows that 
the central cells start to oscillate after all of the cells are excited, 
but this oscillation is not sustained.
In this strong coupling condition,
the boundary cells can not recover their excitability, 
so that the wave front propagating from the center is unable to stop
and reflect to generate successive oscillation. 

Before the oscillation stops, 
there is a narrow parameter region of $2.53\leqslant\tilde{D}\leqslant2.56$, 
where only one side of the ``wall'' alternatively collapses, 
and a complicated period-4 collective oscillation is 
observed (Fig.~\ref{fig:osc_death}(a)).

\subsection{Overall perspective and bifurcation}

\begin{figure}
 \centering
 \includegraphics{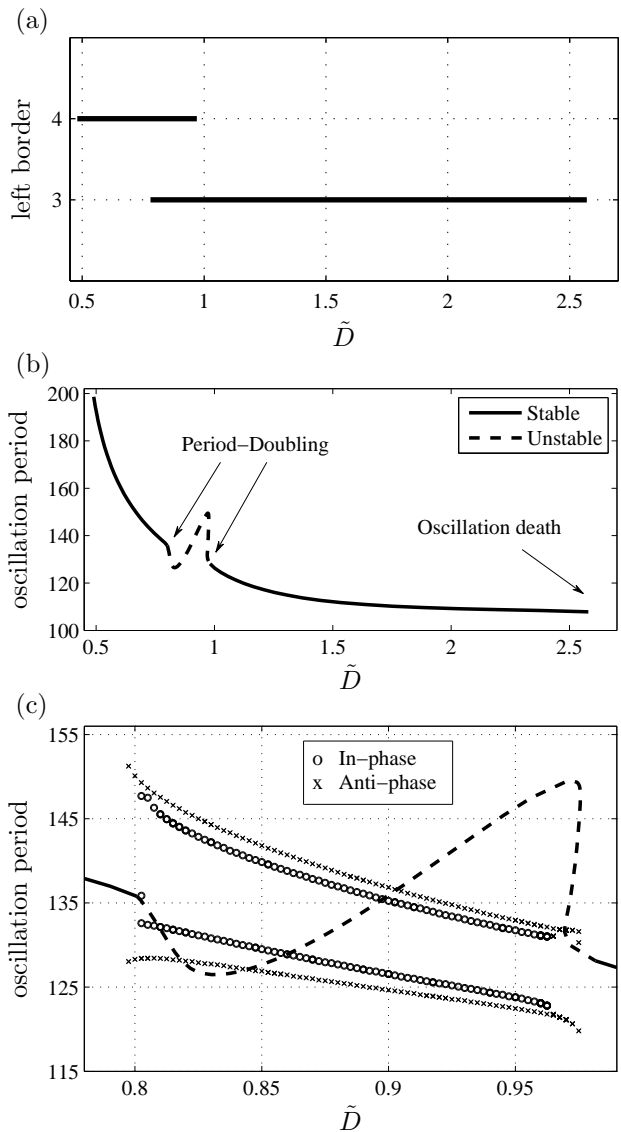}
 \caption{Phase diagram of  
 (a) cell number for the left boundary of collective oscillation.  
 (b) the oscillation period, 
 (c) an enlarged view with in-phase and anti-phase period two solutions, 
 with respect to the change in the strength of coupling.}
 \label{fig:phasediagram}
\end{figure}

There are many factors that may influence the oscillation behavior. 
For example, if the spatial profile of active factor $\Gamma$ becomes 
more ``steep'' rather than a gentle slope, the wave back will tend to  
be locked and fail to reflect from boundary. 
Moreover, if we reduce the excitability of cells by increasing $\epsilon$, 
it will be more difficult for wave front to propagate cross the center, 
and only the half part with stimulation can oscillate. 
Since the global oscillation is induced by mutual coupling, 
we are going to study the oscillation behavior with respect to the 
coupling strength. 
Here, we sweep $\tilde{D}$ from 0.45 to 2.7, 
and summarize the variation in the oscillation period and 
position of the left border of oscillation region. 

If the coupling strength is smaller than 0.48, there is no oscillation, 
and 4 cells from the tissue border are excited while cells 5--16 are silent. 
Oscillation takes place when the wave back passes the 4$^{th}$ cell at 
$\tilde{D}=0.48$. 
The border then shifts between 3 and 4, 
while in-phase and anti-phase period doubled oscillation occur, 
roughly between $0.78<\tilde{D}<0.97$. 
Finally, the oscillation reaches a maximum region: from cells 3 to 18, 
until $\tilde{D}$ is too large for oscillation to occur. 
Figure~\ref{fig:phasediagram}(a) shows the expansion of oscillatory region. 

Variations in the period of oscillation are shown in 
Fig.~\ref{fig:phasediagram}(b).
Once the central cells start to collectively oscillate, 
The period rapidly decreases when the coupling strength increases. 
The rate of the period decrease gradually slows. 
The period changes little in the region where $\tilde{D}$ is large. 
This phenomenon occurs because the stationary interval 
(Fig.~\ref{fig:wave-1p}(e)) greatly contributes to the period of oscillation. 
The decrease in the stationary interval significantly shortens 
the period of oscillation when $\tilde{D}$ is small. 
However, when $\tilde{D}$ is large enough, 
the wave backs reflect immediately without stopping, 
and the period is determined mainly by the velocity of propagation. 
Therefore, the presented oscillation is robust at strong coupling 
condition, and tunable at weak coupling case. 

There is a parameter region (the curve of period is drawn in dashed curve 
Fig.~\ref{fig:phasediagram}(b)) in which system undergoes 
Period-Doubling bifurcation and the period-1 solution lose its stability. 
Meanwhile, period two solutions appear around this region. 
We show more details in Fig.~\ref{fig:phasediagram}(c). 
In the figure, we draw two intervals in a period two solution, 
by measuring the time when $u_4$ positively cross the 
section: $u_4=0.5$. 
Circles ($\circ$) and crosses ($\times$) indicate in-phase and anti-phase 
solutions, respectively. 
The in-phase period-2 solution is the result of Period-Doubing 
bifurcation, while the anti-phase period-2 solution occurs 
saddle-node bifurcation. The anti-phase period-2 solution has 
a wider parameter region than in-phase one, 
and coexistence with fundamental period-1 solution can be observed 
in both side of $\tilde{D}$. 
There are quite complicated bifurcation phenomena specially around 
the occurrence of Period-Doubling bifurcation. 
We have even found period-3 solution (in-phase one around 
$\tilde{D}$=0.802, and anti-phase one around $\tilde{D}=0.975$, respectively). 
Although they are very interesting in the viewpoint of nonlinear dynamical 
system, we leave them to our future work, because current paper 
is going to discuss the possibility of global oscillation and its potential 
applications. 

\section{Discussion}

\subsection{Discreteness vs. continuum}

The above phenomena are observed in a spatially discrete system, 
described by ordinary differential equations. 
As briefly introduced in Sec~\ref{sec:intro},
this discreteness is important in both a mathematical and biological sense.
Let us discuss this significance in more detail.

The diffusion term $D (\partial^2 u / \partial x^2)$
in a one-dimensional spatially continuous reaction-diffusion model
can be formulated as
$D (u_{i-1}+u_{i+1}-2u_i) / \Delta x^2$ in its difference version.
This type of conversion is a common approach to solving PDE numerically.
The diffusion rate $D$ usually does not change much for a specific
substance under constant conditions.
Thus, if we assume that the coupling is mainly 
due to the diffusion-like effects of substances, 
the coupling strength $\tilde{D} \approx D/\Delta x^2$ changes 
in square order with respect to variation of $\Delta x$, 
which biologically corresponds to the distance between cells or the cell size.
Since the profile of the active factor has a scaling property,
it is reasonable to suppose that this gradient works for a field of any size.
Thus, we can study how a change in number of cells $N$ and distant of cells 
$\Delta x$ affects global dynamics.

\begin{figure}[t!]
 \centering
\includegraphics[width=\columnwidth]{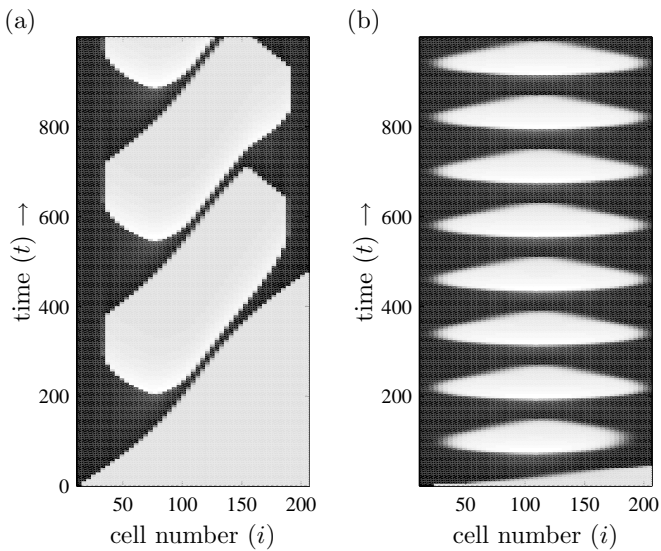}
 \caption{Spatio-temporal diagram of $u_i$.
   Black and white indicate $u_i = 1$ and $u_i = 0$, respectively.
   $N=200$.
   (a) $\tilde{D} = D / \Delta x^2 = 0.7$.
   (b) $\tilde{D} = D / \Delta x^2 = 70$.}
 \label{fig:cell200}
\end{figure}

Figure~\ref{fig:cell200}(a) shows spatio-temporal diagrams 
with a 10-fold increase in the number of cells, $N=200$. 
Other parameters are the same as those in Fig.~\ref{fig:normal_osc}.
Obviously, more time is required for a wave to sweep over the organism.
The period of oscillation and the phase difference 
between the two sides increase greatly.
On the other hand, if the distance $\Delta x$ between cells 
becomes smaller and smaller when cell population increases, 
the system manifests itself more like a continuum
than a discrete system.
In this case, the coupling strength will increase dramatically 
as a square with respect to the decrease in $\Delta x$, i.e., 
a 100-fold increase in $\tilde{D}$ in present case.
When $\tilde{D}$ is as large as 70, 
we have the spatio-temporal diagram
given in Fig.~\ref{fig:cell200}(b).
When we compare this with Fig.~\ref{fig:normal_osc},
there is little change in the period of collective oscillation.
This suggests that the clock tends to run more punctually. 
In a mathematical sense, 
when the population of cells is large enough in a fixed field, 
the behavior of the organism will
follow the solution of a specific partial differential equation, 
which is independent of the number of cells.

Figure.~\ref{fig:cell200}(a) simply corresponds to the case that 
cells grow in an open space, and extend the field by keeping 
size and distant of cells constantly. 
During the initial period of development, however, 
cell divisions within the egg proceed quickly, without much 
increase of the total cell mass and size. 
Thus, cells at this stage rapidly decrease in diameter. 
This may interpret biologically the situation of 
Fig.~\ref{fig:cell200}(b). 

There are many biological situations, however, that the 
intercellular coupling does not follow the diffusion-like ways, 
such as communication involving the delta-notch signaling 
pathway~\cite{Mumm2000151}. 
In those cases, $\Delta x$ has few direct influences on $\tilde{D}$, 
which may represent ``bottlenecks'' irrespective to the diffusion. 
Thus, the modeling based a continuum is inappropriate for some 
conditions.

%

\subsection{Understanding the mechanism}

\begin{figure}[t!]
 \centering
\includegraphics[width=\columnwidth]{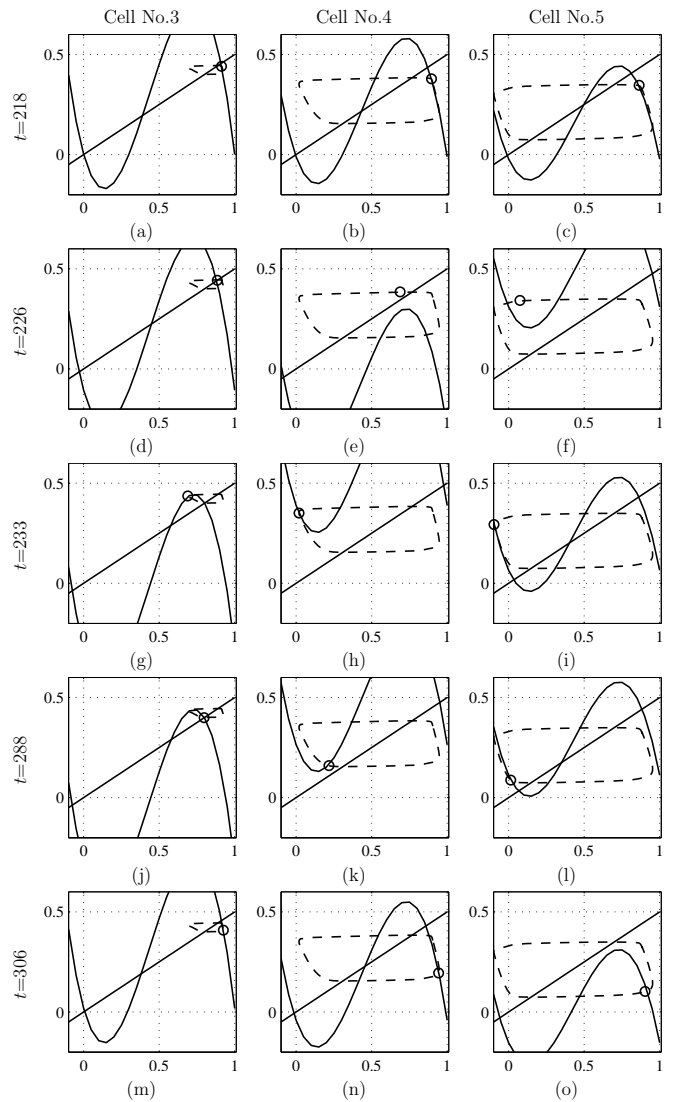}
\caption{Phase portrait diagrams with snapshots of the dynamical
nullcline. Rows indicate the time evolution from the top down,
and columns indicate the number of cells
(3 at left, 4 at middle and 5 at right).
Dashed curves are the limit cycle
solution. Cubic function curves are the nullcline of $\dot{u}_i=0$.
Straight lines are the nullcline of $\dot{v}_i=0$. 
Circles are the position of $(u_i, v_i)$ at specific times.
An animation of the dynamical nullclines can be found at \cite{mysite}.}
\label{fig:nullclinemove}
\end{figure}

Self-sustained collective oscillation is caused by 
the excitability of cells and their mutual interaction.
The system involves complicated bifurcations. 
We present here some qualitative ideas regarding how this oscillation
takes place.

From dynamical equations (\ref{eq:fhn1},\ref{eq:fhn2}) and
their nullcline shown in Fig.~\ref{fig:onoff},
we know that a single cell can exhibit either bistability or mono-stability.
However, if we introduce coupling, 
the geometry of nullclines of one cell will dynamically 
change according to its own state and those of its neighbors.
Because it was assumed that the communication between cells
is only mediated via the activator $u$,
the strict nullcline $\dot{v}_i = 0$ is independent of coupling.
\begin{equation}
\label{eq:nullcline-v0}
v_i = G(u_i) = \beta u_i.
\end{equation}
From Eqs.(\ref{eq:chain1}\&\ref{eq:fhn1}), we obtain the function
for the nullcline $\dot{u}_i=0$ as
\begin{equation}
\label{eq:nullcline-u0}
 v_i = F(u_i,\Gamma_i) =
 \Gamma_i u_i (u_i - \alpha) (1 - u_i) + \Delta U_i,
\end{equation}
where $\Delta U_i = \tilde{D}(u_{i+1}+u_{i-1}-2u_{i})$
is the offset of the cubic function due to coupling.
Thus, the nullcline $v=F(\cdot)$ dynamically moves up and down
in the phase plane, 
corresponding to the state of $u_{i-1}, u_{i}, u_{i+1}$.

In Fig.~\ref{fig:nullclinemove}, we show the phase portrait
of cells around the oscillation border (cells No.3-5),
as well as their dynamical nullcline at some turning points.
Snapshots are taken under the same conditions of normal oscillation
as shown in Fig.~\ref{fig:normal_osc}.

The first row, (a, b, c) of Fig.~\ref{fig:nullclinemove} 
are all in the excited state, 
i.e., for all three cells, $u_i$ is close to 1. 
Therefore, under this condition, 
the vertical offset $\Delta U$ of the nullcline is nearly zero, 
and all three cells exhibit bistability. 
The second row is taken at $t=226$,
when the wave back comes (see Fig.~\ref{fig:wave-1p}(d)),
and the 5$^{th}$ cell moves towards its lower equilibrium 
(Fig.~\ref{fig:nullclinemove}(f)).
Since $\Delta U_4 = \tilde{D}(u_3+u_5-2u_4)$,
a sudden drop in $u_5$ leads to a rapid decrease in the cubic nullcline. 
As shown in Fig.~\ref{fig:nullclinemove}(e),  
the cubic nullcline moves down so that 
the higher equilibrium disappears. 
Thus, the 4$^{th}$ cell becomes mono-stable and the state 
quickly converges to the left branch of the cubic nullcline. 
The decrease in $u_4$ pushes $\Delta U_4$ back to a positive value, 
and makes the nullcline of the 3$^{rd}$ cell move down, 
as shown in Fig.~\ref{fig:nullclinemove}(g, h). 
However, since the 3$^{rd}$ cell has a larger $\Gamma$, 
which controls the amplitude of the cubic nullcline, 
even if $u_4$ decreases to its lowest value 
(Fig.~\ref{fig:nullclinemove}(h)), i.e., 
$\Delta U_3$ reduces to its minimum,  
the cubic and straight nullclines still intersect, 
and the higher equilibrium remain. 
This explains why the wave back passes the 4$^{th}$ cell, 
but stops at the 3$^{rd}$ cell (Fig.~\ref{fig:wave-1p}(a)).
After propagation stops, 
there is a relatively long refractory period from time 230 to 300. 
In this interval, there is a slow decrease in the inhibitor $v_4$. 
Since $u_3 \approx 0.8$ and $u_5 \approx 0$, 
although slight increase in $u_4$ pulls the cubic nullcline down, 
$\Delta U_4$ is still so large that the cubic nullcline 
is above the straight nullcline (Fig.~\ref{fig:nullclinemove}(k)). 
Under this condition, the cell is mono-stable, 
with the equilibrium at the right branch of the cubic function. 
Thus, after a while, the state of $u$ will switch to a  
higher value (Fig.~\ref{fig:nullclinemove}(n)), 
and leads to a reflecting wave front (Fig.~\ref{fig:wave-1p}(f)).
Finally, the states return to the situation of 
Fig.~\ref{fig:nullclinemove}(b) after another refractory period. 

Note that a smaller coupling strength $\tilde{D}$ leads to 
a smaller offset $\Delta U$. 
If we move down the cubic nullcline slightly to cross 
the straight nullcline in Fig.~\ref{fig:nullclinemove}(k), 
the 4$^{th}$ cell becomes bistable. 
This will disable the switch from left to right, 
and stop the oscillation (Fig.~\ref{fig:beforeosc}).

From the above description and Fig.~\ref{fig:nullclinemove}, 
we conclude that the boundary cell, here No.4, 
which is bistable without coupling, 
turns to switch between two types of mono-stable dynamics. 
As introduced in Sec.~\ref{sec:intro}, it is different from the 
studies changing nullclines via coupling to 
an oscillatory geometry. 
This switching becomes the power that underlies the self-sustained 
oscillation observed in the present model. 
The variation of the offset of the dynamical nullcline of the 
boundary cell gives rise to rich oscillation phenomena. 

\subsection{Conditions for oscillation}

We will now explore the conditions for oscillation in an 
approximate manner by studying the dynamics on the oscillation border, 
where wave backs (WB) stop and wave fronts (WF) generate. 
Based on an investigation of the dynamical nullcline and state variable, 
we concluded that a wave back will not pass a critical cell, $c$, 
if the nullclines still intersect at the right branch 
when cell $c+1$ has dropped to 
its lower equilibrium (Fig.~\ref{fig:nullclinemove}(g,h)). 
In contrast, if the intersections disappear, 
the state of $u_c$ will switch to a lower equilibrium. 
Then cell $c$ is possible to oscillate, 
if it fires a wave front, in the case that 
the two nullclines do not cross at the left branch 
when the state of the inhibitor recovers to its lower limit 
(Fig.~\ref{fig:nullclinemove}(e,f)).

Thus, we can roughly solve the condition by finding 
two possible tangency for the two nullclines 
Eq.~(\ref{eq:nullcline-v0}) and Eq.~(\ref{eq:nullcline-u0}). 
This can be achieved using the following equations:
\begin{equation}
\label{eq:tangency}
\frac{{\rm d} F(u, \Gamma)}{{\rm d} u} = \frac{{\rm d} G(u)}{{\rm d} u},
\end{equation}
Equation~(\ref{eq:tangency}) is for two nullclines with the same slope. 
By substituting $F$ and $G$ into Eq.~(\ref{eq:tangency}), we have
\begin{equation}
\label{eq:tangency-expand}
\Gamma(-3  u^2 + 2(1+\alpha)u-\alpha) = \beta,
\end{equation}
from which we obtain two solutions
\begin{equation}
 \label{eq:tangency-solutions}
 u_{T_{1,2}} = \frac{13\pm\sqrt{79-150/\Gamma}}{30}.
\end{equation}

\begin{figure}[t!]
 \centering
 \includegraphics[width=0.9\columnwidth]{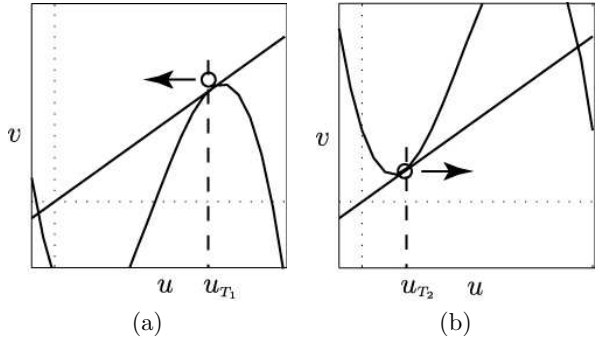}
 \caption{Schematic diagram of two critical tangency situations, 
 corresponding to the conditions for which 
 (a) a wave back passes and (b) a wave front is generated.
 Black circles are the position of states when two nullclines tangent 
 to each other.}
 \label{fig:2con}
\end{figure}

For cell $c$ to propagate a wave back, 
there should be only a lower equilibrium when $u_c$ close 
to the higher tangency point. 
The corresponding condition is $F(u_{T_1}, \Gamma_c) < G(u_{T_1})$. 
Substitution leads to 
\begin{equation}
\label{eq:cond-wb}
 \Gamma_c u_{T_1} (u_{T_1} - \alpha) (1 - u_{T_1}) + 
 \tilde{D} (u_{c-1} + u_{c+1} - 2 u_{T_1}) < \beta u_{T_1}.
\end{equation}
On the other hand, for cell $c$ to generate a wave front, 
there should be no lower equilibrium when $u_c$ is close 
to the lower tangency point. 
This simply means that $F(u_{T_2}, \Gamma_c) > G(u_{T_2})$, 
which can be rewritten as 
\begin{equation}
\label{eq:cond-wf}
 \Gamma_c u_{T_2} (u_{T_2} - \alpha) (1 - u_{T_2}) + 
 \tilde{D} (u_{c-1} + u_{c+1} - 2 u_{T_2}) > \beta u_{T_2}.
\end{equation}
Two critical conditions are shown in Fig.~\ref{fig:2con}.

\noindent \emph{Approximation}

\begin{figure}[t!]
\centering
 \includegraphics{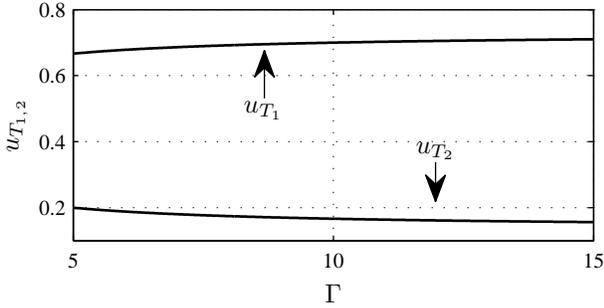}
 \caption{Variation of two tangent points $u_{T_{1,2}}$ with 
 respect to $\Gamma$.}
 \label{fig:ut-vs-G}
\end{figure}

\begin{enumerate}
\item $u_{c-1}$ is the ``distal'' side of the critical cell $c$. 
It remains in its higher equilibrium since the wave back can not pass it. 
Thus, we can approximate it by finding the biggest 
intersection of the two nullclines. 
In the wave back case, since $u_{T_1}$ is close to the higher equilibrium, 
$\Delta U_{c-1}$ is nearly zero. 
Thus, we determine $u_{c-1}$ to be 0.9. 
In the wave front case, however, $u_{T_2}$ is small. 
If we consider the minus $\Delta U_{c-1}$, 
we determine $u_{c-1}$ to be 0.8.
\item $u_{c+1}$ is the ``proximal'' side of the critical cell $c$. 
It switches off before cell $c$ when a wave back comes and waits 
to be excited again by cell $c$, so at the critical time, 
$u_{c+1}$ approaches 0. 
\item Figure~\ref{fig:ut-vs-G} shows $u_{T_{1,2}}$ according to 
Eq.~(\ref{eq:tangency-solutions}). 
Obviously, $u_{T_{1}}$ and $u_{T_2}$ change little, 
and can be regarded as the constants $0.7$ and $0.17$, respectively.
\item Note that the above approximations are not valid 
when the coupling strength is too large, or excitability is too weak. 
\end{enumerate}

\begin{figure}[t!]
\centering
  \includegraphics{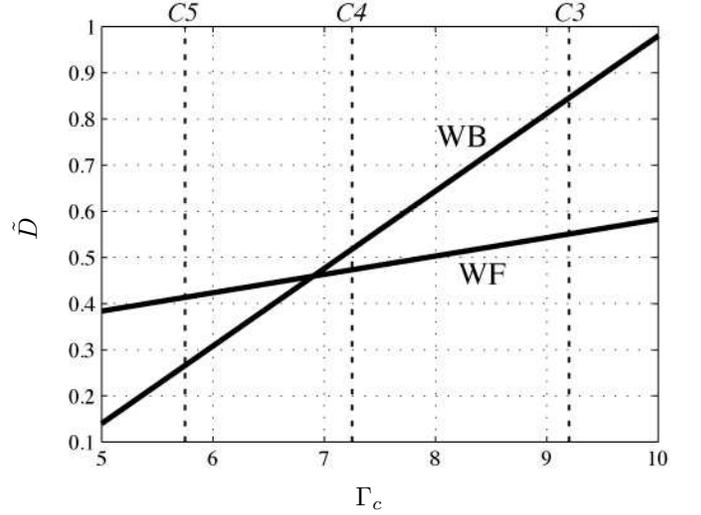}
 \caption{The diagram in the parameter plan ($\Gamma_c, \tilde{D}$) 
representing the conditions of collective oscillation.
A wave back passes the cell, and a wave front reflects, 
when the parameters are above the WB and WF line, 
respectively.}
 \label{fig:D-vs-G-2con}
\end{figure}

According to the above approximations, by substituting 
\[
u_{T_1} = 0.7, u_{c-1} = 0.9, u_{c+1} = 0
\]
into Eq.~(\ref{eq:cond-wb}), and 
\[
u_{T_2} = 0.17, u_{c-1} = 0.8, u_{c+1} = 0
\]
into Eq.~(\ref{eq:cond-wf}), we obtain two rough conditions 
\begin{align}
\label{eq:simp-con-wb}
& \tilde{D} > 0.168 \Gamma_c - 0.7 & \ \ \text{WB passes,}\\
\label{eq:simp-con-wf}
& \tilde{D} > 0.0398 \Gamma_c + 0.1848 & \ \ \text{WF generates.}
\end{align}
Clearly, the critical coupling strength increases linearly 
with respect to the active factor $\Gamma$. 
We draw two lines in Fig.~\ref{fig:D-vs-G-2con}, 
where the WB and WF lines are obtained from 
Eq.~(\ref{eq:simp-con-wb}) and Eq.~(\ref{eq:simp-con-wf}), respectively. 

In Fig.~\ref{fig:D-vs-G-2con}, labels \textit{C3}, \textit{C4} and \textit{C5} 
on the top horizontal axis indicate the value of $\Gamma$ 
defined by Eq.~(\ref{eq:Gamma-profile}) for cells 3, 4 and 5, respectively. 
Lines WF and WB cross each other between \textit{C4} and \textit{C5}. 
This kind of topology makes the 4$^{th}$ and 5$^{th}$ 
cells behave completely different.

For cell 5, WF is above WB. 
If the coupling strength is between WF and WB, 
a wave back coming from the center can switch the 4$^{th}$ cell OFF, 
but the cell can not be switched ON to fire a wave front. 
This is exactly the situation shown in Fig.~\ref{fig:beforeosc}, 
in which no oscillation takes place. 
On the other hand, for cell 4, WF is below WB. 
Clearly, if the coupling strength allows 
the wave back to suppress the 4$^{th}$ cell, 
the cell will be excited again and lead to a wave front. 

The conditions for these two situations depend on many other factors, 
such as the initial conditions, propagation velocity, 
cell excitability, and so on. 
The situation is much more complicated than 
the approximated case we have discussed here. 
Qualitatively, we can conclude that the intersection of 
these two condition lines is the origin of 
self-sustained collective oscillation.  

\section{Concluding remarks}

In this paper,
we have proposed a scenario for self-organized and
self-sustained oscillation in multi-cellular biological tissue.
In contrast to the usual framework based on an oscillatory genetic network,
the present system does not include any self-oscillating cells.  
However, by mutual coupling,  
we can observe collective oscillation inside a group of cells, i.e., tissue. 
Moreover, oscillation can manifests itself in several ways,
corresponding to different coupling strengths.
Anti-phase and in-phase oscillations at the two boundaries
lead to changes in the position of oscillation and the 
oscillating cell population, respectively.
The birth and death of oscillation resulting from variation in the 
coupling strength were also discussed. 
We also provide a general idea of how the size of the cell and 
population affects the oscillatory behavior.
Finally, a detailed investigation of the dynamical movement of 
the nullcline provided insight into the mechanism of complicated 
oscillatory phenomena.  
Although there have been several studies on self-oscillatory
phenomena in spatially discrete systems in the context of
mathematics and physics,
this paper extends these basic ideas to spatio-temporal
self-organization in a biological system.
It is of interest to extent our new hypothesis to spatial three 
dimensional systems, i.e., a more realistic model of living organism. 

Our observations were based on a numerical simulation.
Future analytical studies inspired by 
these interesting phenomenon are needed.
At last, but not less important, 
we are going to cooperate with biologists, 
in order to design corresponding biological experiments and
to explore more proofs supporting our hypothesis.

\end{document}